\documentclass[showkeys,showpacs,superscriptaddress]{revtex4}
\usepackage{amsmath}
\usepackage{amssymb}
\usepackage{graphicx}
\usepackage{float}
\allowdisplaybreaks

\begin{document}

\title{
Dirac/Rarita-Schwinger plus Maxwell theories in $\mathbb{R} \times S^3$ spacetime in the Hopf coordinates
}

\author{
Vladimir Dzhunushaliev
}
\email{v.dzhunushaliev@gmail.com}
\affiliation{
Department of Theoretical and Nuclear Physics,  Al-Farabi Kazakh National University, Almaty 050040, Kazakhstan
}
\affiliation{
Institute of Experimental and Theoretical Physics,  Al-Farabi Kazakh National University, Almaty 050040, Kazakhstan
}
\affiliation{
National Nanotechnology Laboratory of Open Type,  Al-Farabi Kazakh National University, Almaty 050040, Kazakhstan
}

\affiliation{
Academician J.~Jeenbaev Institute of Physics of the NAS of the Kyrgyz Republic, 265 a, Chui Street, Bishkek 720071, Kyrgyzstan
}

\author{Vladimir Folomeev}
\email{vfolomeev@mail.ru}
\affiliation{
Institute of Experimental and Theoretical Physics,  Al-Farabi Kazakh National University, Almaty 050040, Kazakhstan
}
\affiliation{
National Nanotechnology Laboratory of Open Type,  Al-Farabi Kazakh National University, Almaty 050040, Kazakhstan
}

\affiliation{
Academician J.~Jeenbaev Institute of Physics of the NAS of the Kyrgyz Republic, 265 a, Chui Street, Bishkek 720071, Kyrgyzstan
}

\affiliation{
International Laboratory for Theoretical Cosmology, Tomsk State University of Control Systems and Radioelectronics (TUSUR),
Tomsk 634050, Russia
}

\begin{abstract}
We consider the sets of Dirac-Maxwell and Rarita-Schwinger-Maxwell equations in
$\mathbb{R} \times S^3$ spacetime. Using the Hopf coordinates, we show that these equations allow separation of variables and
obtain the corresponding analytic and numerical solutions. It is also demonstrated that the current of the Dirac field is related to the Hopf invariant on the $S^3 \rightarrow S^2$ fibration.
\end{abstract}



\date{\today}

\maketitle

\section{Introduction}

There are many works devoted to solving the Dirac equation. The latter has different types of solutions both in the case of flat spacetime
(see, e.g., the textbook~\cite{Greiner:1990tz}) and in curved background spacetime within general relativity (see, e.g., the textbook~\cite{Chandrasekhar}).
In the present study, we will be interested in solving the Dirac equation in spacetime with a spatial cross-section in the form of a three-dimensional sphere $S^3$.
An interesting feature of such a solution is that it is related quite nontrivially to the Hopf invariant on such a sphere. Recall that
this invariant arises when considering a sphere $S^3$ as the fibration $S^3 \rightarrow S^2$ with fibres $S^1$. Such a fibration has a topological invariant~--
the Hopf invariant. For this reason, we will obtain a topologically nontrivial solution, unlike the topologically trivial one on the sphere
 $S^3$ found in Ref.~\cite{Goatham:2009ek}.

For the Rarita-Schwinger equation, much fewer solutions are known. Ref.~\cite{Gueven:1980be}
studies equations for a spinor field of a spin of  $3/2$ on the background of the Kerr-Reissner-Nordstr\"{o}m spacetime, and it is shown that the equations allow separation of variables.
In Ref.~\cite{Gueven:1982tk},
the spin-3/2 perturbations of the extreme Reissner-Nordstr\"{o}m black holes are analysed. In Ref.~\cite{TorresdelCastillo:1990aw}, the spin-3/2 perturbations
of a rotating black hole are analysed within the framework of linearised supergravity.
A spin-3/2 equation for a massless field is solved on the Kerr-Newman black-hole background in Ref.~\cite{Anton:1988cz}. Special solutions of the Rarita-Schwinger
equation in spacetimes admitting shear free congruences of null geodesics are found in Ref.~\cite{Szereszewski:2002sx}. The general Lagrangian and propagator
for a vector-spinor field in $d-$dimensions are obtained in Ref.~\cite{Pilling:2004cu}. The propagation and quantization of Rarita-Schwinger waves in an external
electromagnetic potential have been considered in Ref.~\cite{Velo:1969bt}.

The Rarita-Schwinger equation is one of fundamental equations in supergravity (SUGRA),
which in turn is one of candidates for the role of the quantum gravity theory. In SUGRA,
there are many corresponding solutions in the bosonic sector. For example, Ref.~\cite{Pang:2015vna} considers the theory containing two
scalar fields and two pseudoscalar fields; in Ref.~\cite{Hoare:2015wia}, an exact type IIB supergravity solution that represents a one-parameter deformation of
the $T$-dual of the $AdS_5 \times S^5$ background was found; Ref.~\cite{Gibbons:2008vi} is devoted to finding a general fully nonlinear solution
of topologically massive supergravity admitting a Killing spinor. To the best of our knowledge, there are no solutions supported by spinor fields in SUGRA
(this applies, in particular, to a Rarita-Schwinger field). Therefore, the derivation of solutions to the Rarita-Schwinger equation, even in the absence of gravity, is of great interest.

The purpose of the present study is to obtain the following solutions in $\mathbb{R} \times S^3$ spacetime:
(a)~topologically nontrivial solutions of the Dirac equation;
(b)~solutions to the Rarita-Schwinger equation; and (c)~self-consistent solutions within Dirac-Maxwell and Rarita-Schwinger-Maxwell theories.

The paper is organized as follows. In Sec.~\ref{DM}, we consider Dirac-Maxwell theory, within which we obtain an analytic solution without electromagnetic field and a numerical solution
with electromagnetic field. In Sec.~\ref{RSM}, we consider Rarita-Schwinger-Maxwell theory, within which we find an analytic solution in the absence of electromagnetic field.

\section{Dirac-Maxwell theory}
\label{DM}

\subsection{$\mathfrak{Ansatz}$ and equations}

We consider Dirac-Maxwell theory, where the source of electromagnetic field is taken in the form of a massless Dirac field. The corresponding Lagrangian
can be written as (hereafter, we work in units where $\hbar=c=1$)
\begin{equation}
	L =	\frac{i} {2} \left(
			\bar \psi \gamma^\mu \psi_{; \mu} -
			\bar \psi_{; \mu} \gamma^\mu \psi
		\right) - \frac{1}{4} F_{\mu\nu}F^{\mu\nu}
\label{lagr_tot}
\end{equation}
which contains the covariant derivative
$
\psi_{; \mu} \equiv \left[\partial_{ \mu} +
1/8\, \omega_{a b \mu}\left(
	\gamma^a  \gamma^b -
	\gamma^b  \gamma^a\right) -
	i e A_\mu
\right]\psi
$,
where $\gamma^a$ are the Dirac matrices in
flat space (below we use the spinor representation of the matrices);
$a, b$ and $\mu, \nu$ are tetrad and spacetime indices, respectively;
$
	F_{\mu \nu} = \partial_\mu A_\nu - \partial_\nu A_\mu
$ is the electromagnetic field tensor; $A_\mu$ are four-potentials of the electromagnetic field; $e$ is a charge in Maxwell theory.
In turn, the Dirac matrices in curved space,
$\gamma^\mu = e_a^{\phantom{a} \mu} \gamma^a$, are obtained using
the tetrad
$ e_a^{\phantom{a} \mu}$, and $\omega_{a b \mu}$ is the spin connection
[for its definition, see Ref.~\cite{Lawrie2002}, formula (7.135)].

Varying the corresponding action with the Lagrangian \eqref{lagr_tot}, one can obtain the following set of equations:
\begin{eqnarray}
	i \gamma^\mu \psi_{;\mu} &=& 0 ,
\label{D_10}\\
	\frac{1}{\sqrt{-g}}
	\frac{\partial }{\partial x^\nu}
	\left( \sqrt{-g} F^{\mu \nu} \right) &=& - 4 \pi j^\mu,
\label{D_15}
\end{eqnarray}
where $g$ is the determinant of the metric tensor and $j^\mu=e \bar\psi\gamma^\mu\psi$ is the four-current.

We seek a solution of the above equations in $\mathbb{R} \times S^3$ spacetime with the Hopf coordinates $\chi, \theta, \varphi$ on a sphere~$S^3$.
The corresponding metric is
\begin{equation}
	ds^2 = dt^2 - \frac{r^2}{4}
	\left[
		\left( d \chi - \cos\theta d \varphi\right)^2
		+ d \theta^2 + \sin^2 \theta d \varphi^2
	\right] = dt^2 - r^2 dS^2_3 ,
\label{D_20}
\end{equation}
where $dS^2_3$ is the Hopf metric on the unit $S^3$ sphere; $r$ is a constant;
$0\leq \chi, \varphi \leq 2\pi$ and $0\leq \theta \leq \pi$.
For a detailed description of the Hopf fibration, see Appendix~\ref{Hopf_bundle}.

To solve Eqs.~\eqref{D_10} and \eqref{D_15}, we use the following
$\mathfrak{Ansatz}$ for the spinor and electromagnetic fields:
\begin{align}
	\psi_{mn} &= e^{-i  \Omega t} e^{i n \chi} e^{i m \varphi}
	\begin{pmatrix}
		\Theta_1(\theta)  \\
		\Theta_2(\theta)  \\
		0  \\
		0
	\end{pmatrix} ,
\label{D_30}\\
	A_\mu &= \left\lbrace
		\phi(\theta), r \rho(\theta) ,0, r \lambda(\theta)
	\right\rbrace  .
\label{D_35}
\end{align}
where $m,n$ are integers. The spinor may transform under a rotation through an angle $2 \pi$  as
$$
	\psi_{mn}(\chi + 2 \pi, \theta, \varphi + 2 \pi) =
	\psi_{mn}(\chi , \theta, \varphi).
$$
Then, due to the presence of the  factors $e^{i n \chi}$ and $e^{i m \varphi}$ in Eq.~\eqref{D_30},
the spinors $\psi_{mn}$ and $\psi_{pq}$ with different pairs of indices  $(m,n)$ and $(p, q)$ are  orthogonal.

To solve the equations, we employ the tetrad
$$
  e^a_{\phantom{a} \mu} =
  \begin{pmatrix}
    1 & 		0 		& 0 			& 0 						\\
    0 & 	\frac{r}{2}	& 0 			& - \frac{r}{2} \cos \theta 	\\
    0 & 0				& \frac{r}{2} 	& 0 						\\
    0 & 0				& 0 			& \frac{r}{2} \sin \theta	
  \end{pmatrix}
$$
coming from the metric \eqref{D_20}.
Taking all this into account,
the current $j^\mu$ [for its definition, see after Eq.~\eqref{D_15}] is
$$
	j^\mu = 	- \frac{e}{r}
	\left\lbrace
		r \left( \Theta_1^2 + \Theta_2^2 \right) ,
		- 2 \left[ \cot \theta \left( \Theta_1^2 - \Theta_2^2\right)  +
		2 \Theta_1 \Theta_2
		\right] ,
		0 ,
		- 2 \csc \theta
		\left(
			\Theta_1^2 - \Theta_2^2
		\right)
	\right\rbrace.
$$
After substitution of the $\mathfrak{Ansatz}$ \eqref{D_30} and \eqref{D_35} in Eqs.~\eqref{D_10} and \eqref{D_15},
we have
\begin{align}
	& \Theta_{1}^\prime + \Theta_{1} \left(
		\frac{\cot \theta}{2} + n + e r \rho
	\right) +
	\Theta_{2} \left(
		\frac{1}{4} - \frac{r \Omega}{2} - n \cot \theta -
		\frac{m}{\sin \theta} - e r \rho \cot \theta -
		\frac{e r \lambda}{\sin \theta}	+ \frac{e r}{2} \phi
	\right) = 0 ,
\label{D_50}\\
	&\Theta_{2}^\prime + \Theta_{2} \left(
		\frac{\cot \theta}{2} - n - e r \rho
	\right) + \Theta_{1}  \left(
		- \frac{1}{4} + \frac{r \Omega}{2} - n \cot \theta -
		\frac{m}{\sin	\theta}  - e r \rho \cot \theta -
				\frac{e r \lambda}{\sin \theta} -
				\frac{e r}{2} \phi
	\right) = 0 ,
\label{D_60}\\
	& \frac{1}{\sin \theta} \left(
		\sin \theta \phi^\prime
	\right)^\prime = \phi^{\prime \prime} +
	\cot \theta \phi^ \prime =
	- \frac{e r^2}{4} \left(
		\Theta_1^2 + \Theta_2^2
	\right) ,
\label{D_70}\\
	&\left(
		\frac{\rho^\prime}{\sin \theta}
	\right)^\prime +
	\left(
		\cot \theta \lambda^\prime
	\right)^\prime =
	\frac{\rho^{\prime \prime}}{\sin \theta} -
	\frac{\cot \theta}{\sin \theta} \rho^\prime +
	\cot \theta \lambda^{\prime \prime} -
	\frac{\lambda^\prime }{\sin^2 \theta} =
	\frac{e r^2}{8} \left[
		\cos \theta \left( \Theta_1^2 - \Theta_2^2 \right) +
		2 \sin \theta \Theta_1 \Theta_2
	\right] ,
\label{D_80}\\
	&\left(
		\frac{\lambda^\prime}{\sin \theta}
	\right)^\prime +
	\left(
		\cot \theta \rho^\prime
	\right)^\prime =
		\frac{\lambda^{\prime \prime}}{\sin \theta} -
		\frac{\cot \theta}{\sin \theta} \lambda^\prime +
		\cot \theta \rho^{\prime \prime} -
		\frac{\rho^\prime }{\sin^2 \theta} =
	\frac{e r^2}{8}
		\left( \Theta_1^2 - \Theta_2^2 \right) ,
\label{D_90}
\end{align}
where the prime denotes differentiation with respect to $\theta$. Note that the set of equations
\eqref{D_50}-\eqref{D_90} must be solved as an eigenvalue problem for $\Omega$
with the eigenfunctions $\Theta_{1, 2}$.

We seek a solution in the following particular form:
$$
	\Theta_2 = \pm \Theta_1 = \Theta , \quad \lambda^\prime = - \rho^\prime \cos \theta ,
$$
and the eigenparameter $\Omega$ is given by the expression
\begin{equation}
	r \Omega_n = \frac{1}{2} \pm 2 n .
\label{D_93}
\end{equation}
It is seen from this expression that the energy of the system $\Omega$ depends on $n$ only, i.e., the energy spectrum is degenerate with respect to $m$. 

For the above particular solution, Eq.~\eqref{D_90}
is satisfied identically, and the remaining equations yield
the set of equations for the functions $\Theta, \phi,  \rho$,  and $\lambda$,
\begin{align}
	\tilde \Theta^\prime \mp \tilde \Theta
	\left[
		\cot \theta \left(\mp \frac{1}{2} + n +  \tilde \rho\right) + \frac{1}{\sin \theta}\left(m + \tilde  \lambda\right)
	\right] &= 0 ,
\label{D_140}\\
	\frac{1}{\sin \theta} \left(
			\sin \theta \tilde \phi^\prime
		\right)^\prime &=
		- \frac{\tilde \Theta^2}{2}  ,
\label{D_150}\\
	\frac{1}{\sin \theta} \left(
		\sin \theta {\tilde\rho}^\prime
	\right)^\prime &= \pm \frac{\tilde \Theta^2}{4}  ,
\label{D_160}\\
	\tilde \lambda^\prime &= -  \tilde \rho^\prime \cos \theta ,
\label{D_170}
\end{align}
where we have introduced new dimensionless variables $\tilde \rho, \tilde \lambda, \tilde \phi=e r \left(\rho, \lambda, \phi\right)$ and $\tilde \Theta =e r^{3/2}\Theta$.
From the Maxwell equations \eqref{D_150} and \eqref{D_160}, it then follows that $\tilde \phi = \mp 2  \tilde \rho$.

\subsection{Analytic solutions with ``frozen'' electric and magnetic fields
}

Before solving the full set of equations \eqref{D_140}-\eqref{D_170}, consider the case of  ``frozen'' electric and magnetic fields, where
the scalar and vector potentials $\phi = \rho = \lambda = 0$.
The solution of Eq.~\eqref{D_140} is then \cite{Dzhunushaliev:2020dom}
\begin{equation}
\begin{split}
	\left( \Theta_{1} \right)_{mn} &=
	\pm \left( \Theta_{2} \right)_{mn} = \pm \Theta_{mn} =
	\frac{C}{r^{3/2}} \sin^\alpha \left( \frac{\theta}{2} \right)
	\cos^\beta \left( \frac{\theta}{2} \right) ,
\\
	 \alpha &= \pm\left(n+m\right) - \frac{1}{2}, \quad
		\beta = \pm \left(n-m\right) - \frac{1}{2} ,
	\quad
		\Omega_n = \frac{1}{2 r}\left(1\pm 4 n\right).
\label{3_a_1_30}
\end{split}
\end{equation}
Here, we explicitly take the integration constant in the form  $C/r^{3/2}$ so that the constant $C$ would be dimensionless.

Since the spinors $\psi_{mn}$ and $\psi_{pq}$ are orthogonal for $(m,n) \neq (p, q)$, 
using the normalization condition
$$
	\int \psi^\dagger_{mn} \psi_{mn} \sin \theta d \chi d \theta d \varphi =
	\frac{4 C^2}{r^3} \left( 2 \pi \right)^2
	\frac{
		\Gamma \left( n + m + \frac{1}{2}\right)
		\Gamma \left( n - m + \frac{1}{2}\right)
	}{\Gamma \left( 2 n + 1 \right) } = 1 ,
$$
one can find the normalization constant
$$
	C_{mn} = \frac{1}{4 \pi} \sqrt{r^3
		\frac{\Gamma \left( 2 n + 1 \right)}{
			\Gamma \left( n + m + \frac{1}{2}\right)
			\Gamma \left( n - m + \frac{1}{2}\right)
		}
	} .
$$

The condition of the positiveness of the $\Gamma$ functions leads to the following inequalities for the quantum numbers $m$ and $n$:
$$
	n \geqslant 0 , \left| m \right| \leqslant n .
$$
In this case, the solution energy spectrum \eqref{D_93} satisfies the inequalities
$$
	r \Omega_{+,n} = \frac{1}{2} + 2 n \geqslant \frac{1}{2} ,
	\quad
	r \Omega_{-,n} = \frac{1}{2} - 2 n \leqslant \frac{1}{2} .
$$

Consider now some physical properties  of the solution~\eqref{3_a_1_30}.
The spin density tensor for the Dirac field is defined as
$$
	S_{abc} =  \frac{1}{8} \bar{\psi} \left\lbrace
		\gamma_a , \sigma_{bc}
	\right\rbrace \psi ,
$$
where $\sigma_{ab} = (i/2) \left[
		\gamma_a , \gamma_b
	\right]$.
After substituting the $\mathfrak{Ansatz}$ \eqref{D_30}, we have the following nonzero components of this tensor:
$$
	S_{012} = - \frac{\mathcal{D}}{4} ,\quad
	S_{023} = \frac{\mathcal{C}}{2} ,\quad
	S_{123} = - \frac{\mathcal{A}}{4} ,
$$
where
$
 \mathcal{A} = \Theta_1^2 + \Theta^2_2 ,
	\mathcal{C} = \Theta_1 \Theta_2 ,
	\mathcal{D} = \Theta_1^2 - \Theta^2_2
$.
After substituting the solution~\eqref{3_a_1_30}, we get the absolutely antisymmetric tensor
$$
	S_{023} = - S_{123} =
	\frac{C^2}{2}
	\sin^{2 m + 2 n - 1}\left(\frac{\theta }{2}\right)
	\cos^{- 2 m + 2 n - 1}\left(\frac{\theta }{2}\right) .
$$

The current $j^a$ is defined as
$$
	j^a =
	\left(
		\Theta_1^2 + \Theta^2_2 ,
		2 \Theta_1 \Theta_2 , 0,
		\Theta_1^2 - \Theta^2_2
	\right) ,
$$
and for the exact solution~\eqref{3_a_1_30} it is
$$
	j^a =
	2 \Theta^2
	\left(
		\omega^0 + \omega^1
	\right)
$$
with the 1-forms
$
	\omega^0 = dt, \;
	\omega^1 = d \chi - \cos\theta d \varphi .
$

The 1-form $\omega^1$ defines the Hopf invariant (see Appendix~\ref{Hopf_bundle})
$$
	H = \frac{1}{V} \int \omega^1 \wedge d \omega^1 = 1.
$$

The energy-momentum tensor is
\begin{equation}
	T_{ab} = \frac{1}{r}
	\begin{pmatrix}
		(2 n + \frac{1}{2}) \mathcal{A} &
		- n \mathcal{B}	&	0	&	
		- \left( n  + \frac{1}{2}\right) \mathcal{D} -
		\left( \frac{m}{\sin \theta} + n \cot \theta \right) \mathcal{A} \\
			- n \mathcal{B}	&	- \frac{\mathcal{A}}{2} + 4 n \mathcal{C}	&	0	&
			n \mathcal{B} - \frac{\cot \theta }{2} \mathcal{A} +
			\frac{2 m}{\sin \theta} \mathcal{C} \\
			0	&	0	&
			\frac{\mathcal{A}}{2} + 2 \mathcal{E} &	0	\\
			- \left( n  + \frac{1}{2}\right) \mathcal{D} -
			\left( \frac{m}{\sin \theta} + n \cot \theta \right) \mathcal{A}	&
			n \mathcal{B} - \frac{\cot \theta }{2} \mathcal{A} +
			\frac{2 m}{\sin \theta} \mathcal{C}	&	0	&	
			\frac{\mathcal{A}}{2} +
			2 \left(
				\frac{m}{\sin \theta} + n \cot \theta
			\right) \mathcal{D}
	\end{pmatrix} ,
\label{2_c_55}
\end{equation}
where $\mathcal{B} = \left( \Theta_1 + \Theta_2\right)^2$ and $\mathcal{E} = \Theta_1^\prime \Theta_2 - \Theta_2^\prime \Theta_1$.
For the solution~\eqref{3_a_1_30}, we have
\begin{equation}
\begin{split}
	T_{ab} =&
	\frac{2 c^2 }{r}
	\sin^{2 m + 2 n - 1}\left(\frac{\theta }{2}\right)
	\cos^{- 2 m + 2 n -1}\left(\frac{\theta }{2}\right)
	\nonumber\\
	&
\times	\left(
	\begin{array}{cccc}
		 2 n + \frac{1}{2} & - 2 n & 0 &
		 - \frac{m}{\sin \theta} -\left( n - \frac{1}{2} \cot \theta \right)  \\
		 - 2 n & 2 n - \frac{1}{2} & 0 &
		 \frac{m}{\sin \theta} + \left( n - \frac{1}{2} \cot \theta \right) \\
		 0 & 0 & 1 & 0 \\
		 - \frac{m}{\sin \theta} -\left( n - \frac{1}{2} \cot \theta \right) &
		 \frac{m}{\sin \theta} + \left( n - \frac{1}{2} \cot \theta \right) & 0 & 1 \\
	\end{array}
	\right) .
\end{split}
\end{equation}


\subsection{Numerical solution with electric and magnetic fields
}
\label{numerical}

Since Eqs.~\eqref{D_140}-\eqref{D_170} cannot be integrated analytically, we will seek their numerical solution. Before doing
that, it is necessary to find an approximate solution near the point $\theta = 0$,
which may be sought in the form of the following power series:
\begin{equation}
	\tilde \Theta = \theta^{\alpha } \sum _{i=0} a_i \theta^i , \quad
\tilde \phi = \theta^{\gamma } \sum _{i=0} f_i \theta^i , \quad
\tilde \lambda = \theta^{\delta } \sum _{i=0} c_i \theta^i , \quad
\tilde \rho = \theta^{\epsilon } \sum _{i=0} d_i \theta^i ,
\label{3_e_10}
\end{equation}
where $\alpha, \gamma, \delta, \epsilon, a_i, f_i, c_i, d_i$ are parameters to be determined.
Substituting \eqref{3_e_10} in Eqs.~\eqref{D_140}-\eqref{D_170}
and comparing the leading terms, we find the values of the expansion parameters
$$
	\alpha = m + n - \frac{1}{2} , \quad
	\epsilon = \delta = 2 \alpha + 2 , \quad
	-d_0  = c_0 = - \frac{a_0^2}
	{16 \left(	m + n + \frac{1}{2}	\right)^2} , \quad
	a_1  = d_1 = - c_1 = 0 .
$$
Thus, for numerical computations,
we have the following values of the functions at the point $\theta = \Delta$:
$$
		\tilde \Theta(\Delta) = a_0 \Delta^{m + n - 1/2} , \quad
	\tilde \lambda(\Delta) = - \frac{a_0^2}{
			16 \left(
				m + n + \frac{1}{2}
			\right)^2
			} \Delta^{2(m + n) + 1} , \quad
	\tilde \rho(\Delta) = \frac{a_0^2}{
				16 \left(
					m + n + \frac{1}{2}
				\right)^2
				} \Delta^{2(m + n) + 1} ,
$$
where the parameter $a_0$ is arbitrary.
Since the functions $\tilde \rho$ and $\tilde \phi$ are linearly proportional [see the corresponding expression after Eq.~\eqref{D_170}], for $\tilde \phi$,
there will be an expansion similar to that given above for $\tilde \rho$.
Also, since the set of equations~\eqref{D_140}-\eqref{D_170}  contains $\sin\theta$ in the denominator,
 we start the numerical solution not from the point $\theta = 0$ but from $\theta=\Delta \ll 1$.

In the case under consideration, the components of the electric, $E_\theta$, and magnetic, $H_{\chi, \varphi}$,
fields are
\begin{align}
	E_\theta &= F_{t \theta} = \phi^\prime , \quad
	H_\chi = \frac{1}{2} \epsilon_{\chi jk} F^{jk} =
	2 \frac{\rho^\prime \cos \theta + {\lambda^\prime}}{\sin \theta}
	= 0 ,
\nonumber\\
	H_\varphi &= \frac{1}{2} \epsilon_{\varphi jk} F^{jk} =
	-	2 \frac{\rho^\prime + \lambda^\prime \cos \theta}{\sin \theta} =
	-2 \rho^\prime \sin \theta =
	\pm \phi^\prime \sin \theta .
\nonumber
\end{align}

\begin{figure}[t]
\begin{minipage}[t]{.45\linewidth}
\centering{
	\includegraphics[width=1\linewidth]{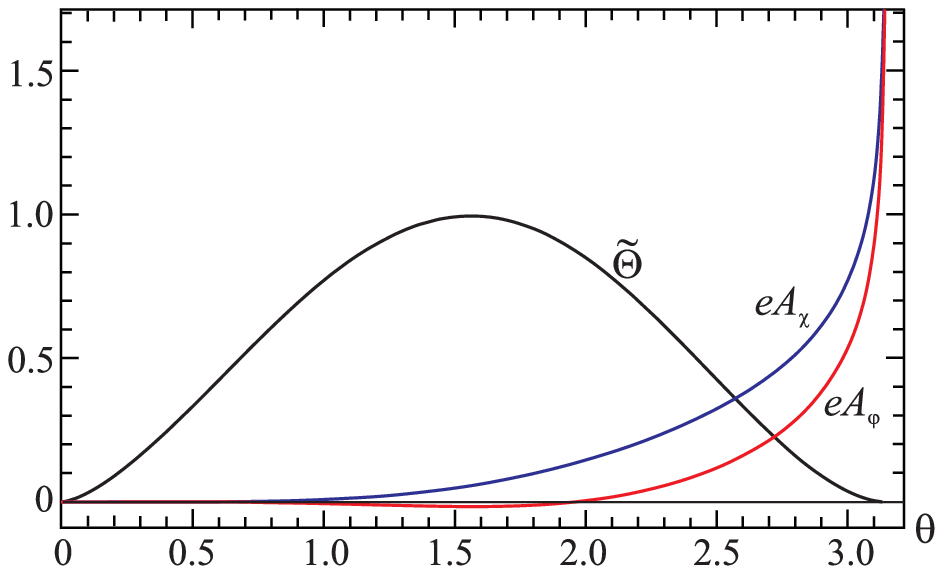}
}
\vspace{-0.5cm}
\caption{
	The spinor $\tilde\Theta$ and the electromagnetic field potentials $e A_{\chi}=\tilde \rho$ and  $e A_{\varphi}=\tilde\lambda$
for the case of $m = 0, n = 2$, and $a_0 = 1$.
}
\label{functions}
\end{minipage}\hfill
\begin{minipage}[t]{.45\linewidth}
\centering{
	\includegraphics[width=1\linewidth]{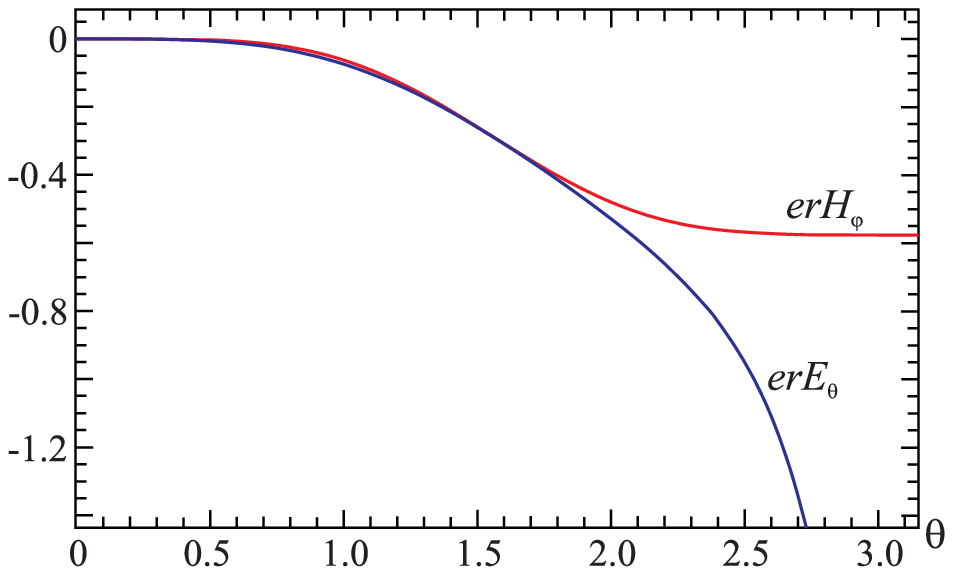}
}
\vspace{-0.5cm}
\caption{
	The electric, $e r E_\theta=\tilde\phi^\prime$, and magnetic,  $e r H_{\varphi}=-2\tilde\rho^\prime\sin\theta$, fields for the case of
	$m = 0, n = 2$, and $a_0 = 1$.
}
\label{fields}
\end{minipage}\hfill
\end{figure}

The results of calculations are given in Figs.~\ref{functions} and \ref{fields} 
[note that for plotting these graphs we have chosen the upper signs in Eqs.~\eqref{D_140} and \eqref{D_160}].
The analysis of this numerical solution indicates that the function
$\Theta$ is practically coincide with the analytic solution \eqref{3_a_1_30} obtained in the absence of the fields $\phi, \lambda$, and $\rho$.
In turn, using the solutions found, one can also calculate the energy-momentum tensor~\eqref{2_c_55}.

\subsubsection{Solution in the neighborhood of singular points
}

It is seen from the graphs shown in Fig.~\ref{functions} that as $\theta\to \pi$
the functions $\rho$ and $\lambda$ diverge. In this connection, it is useful to analyse their behavior near the point $\theta= \pi$,
where a solution is sought in the form
$$
	\tilde \Theta = \tilde \theta^{\alpha } \sum _{i=0} a_i \tilde \theta^i , \quad
	\tilde \lambda = \lambda_0 \ln \tilde \theta +
	\mathcal O \left( \tilde \theta^{\delta }\right) , \quad
	\tilde \rho = \rho_0 \ln \tilde \theta +
		\mathcal O \left( \tilde \theta^{\epsilon }\right)
$$
with $\tilde \theta = \pi - \theta$. Making a change of the variable $\theta \to \pi - \tilde \theta$,
one can find from Eq.~\eqref{D_170} that as $\theta\to \pi$ the parameter $\rho_0 = \lambda_0$,
and Eq.~\eqref{D_160} is satisfied to within $\tilde \theta^{2 \alpha }$. Then the terms with
$\rho_0 \ln \tilde \theta$ and $\lambda_0 \ln \tilde \theta$ in the square brackets of Eq.~\eqref{D_140}
cancel each other. That will leave the term $\left(\mp 1/2+n+m\right)\tilde\Theta/\sin\tilde\theta $, whose behavior
is determined by the behavior of $\tilde\Theta$, which in turn depends on the values of the parameters  $m$ and $n$.
In particular, for the solutions with $m=0$  and $n=2$ shown in Fig.~\ref{functions} this term tends to zero.
In turn, at the point  $\theta = \pi$, the electromagnetic potentials
 $\tilde \lambda$ and $\tilde \rho $ and the component of the electric field $E_\theta \propto \tilde \rho^\prime $ diverge,
 whereas the magnetic field
$
	H_\varphi \propto \tilde \rho^\prime \sin \tilde \theta
	\to \text{const}
$ (see Fig.~\ref{fields}).

Note also that we can start the solution not from the point $\theta=0$ but from $\theta=\pi$.
Then, by changing to the new variable  $\tilde \theta = \pi - \theta$ and seeking a solution of
Eqs.~\eqref{D_140}-\eqref{D_170} in the same form~\eqref{3_e_10},
by comparing the leading terms, we get the following values of the expansion coefficients:
$$
	\alpha = m + n + \frac{1}{2} , \quad
	\epsilon = \delta = 2 \alpha + 2 , \quad
	d_0  = c_0 = \frac{a_0^2}
	{16 \left(	m + n + \frac{3}{2}	\right)^2} , \quad
	a_1  = d_1 = c_1 = 0 .
$$
In this case the solution will be regular at the point $\theta=\pi$ but singular for $\theta=0$.
Then the corresponding graphs of the solutions will be mirror-inverted compared to those shown
in Figs.~\ref{functions} and \ref{fields}.

\subsubsection{Energy-momentum tensor of the electromagnetic field }

Let us now write out the tetrad components of the energy-momentum tensor of the electromagnetic field for the potential~\eqref{D_35}.
Using Eq.~\eqref{D_170}, one can find
$$
 T_{a b} = \frac{8}{r^2}
	\begin{Bmatrix}
			\frac{1}{4}\left(4\rho^{\prime 2}+\phi^{\prime 2}\right) &	\phi^\prime \rho^\prime	&	0	&	0	\\	
	\phi^\prime \rho^\prime	&	\frac{1}{4}\left(4\rho^{\prime 2}+\phi^{\prime 2}\right) &	0	&		0 \\
	0	&	0	&	\rho^{\prime 2}-\frac{1}{4}\phi^{\prime 2}	 &	0	\\
	0&	0	&	0	&	-\rho^{\prime 2}+\frac{1}{4}\phi^{\prime 2}
		\end{Bmatrix}.
$$
Taking into account that the potentials  $\rho$ and $\phi$ appearing here are related by $\phi=\mp 2\rho$
[see after Eq.~\eqref{D_170}],  $T_{tt}$,  $T_{\chi\chi}$, and $T_{t\chi}$ components diverge at the point $\theta = \pi$, and the components
$T_{\theta\theta}$ and $T_{\varphi\varphi}$ are equal to zero.

It is seen from the above results that we are dealing with a very interesting situation:
(a)~ the set of equations \eqref{D_140}-\eqref{D_170} has the regular solution for the spinor function
 $\Theta$ and singular solutions for the electromagnetic functions $\rho$ and $\lambda$;
 (b)~either at the point $\theta=\pi$ or for $\theta=0$ the behavior of the potentials $\rho$ and $\lambda$ is such that the electric field
 $E_\theta$ is singular and the magnetic field $H_\varphi$ is regular; and
 (c)~the energy density and the pressure of the electromagnetic field, as well as the
    $(t-\chi)$ component of the Poynting vector, are singular either at the point $\theta=\pi$ or for  $\theta=0$.

\subsubsection{Small electric and magnetic fields
}

Consider the case of small electric and magnetic fields, where one can neglect the terms with
$e r$ in Eqs.~\eqref{D_50}-\eqref{D_90} compared to other terms. As a result, there will be the set of Maxwell's equations~\eqref{D_150}-\eqref{D_170}
plus the separate Dirac equation~\eqref{D_140}, whose solution is given by Eq.~\eqref{3_a_1_30}.

A first integration of~\eqref{D_150} yields
$$
	\phi^\prime = - \frac{e C^2}{2 r}
	\frac{B_{\sin(\theta/2)}(1 + \alpha, 1 + \beta)}{\sin \theta} ,
	\quad \alpha > -1,
$$
where $B_{\sin(\theta/2)}(1 + \alpha, 1 + \beta)$ is the incomplete Euler beta function.
Using the expansion in the neighbourhood of $\theta =0$,
$$
	\frac{B_{\sin(\theta/2)}(1 + \alpha, 1 + \beta)}{\sin \theta}
	\approx \frac{\theta^{2 \alpha + 1}}{4^{1 + \alpha} (1 + \alpha)} ,
$$
we have for the electric and magnetic field intensities
$$	
\left| E_\theta \right| \approx  \theta^{2 \alpha + 1} , \quad
	\left| H_\varphi \right| \approx  \theta^{2 \alpha} .
$$
This means that for $\alpha > 0$  these fields go to zero as $\theta \rightarrow 0$.

\section{Rarita-Schwinger-Maxwell theory}
\label{RSM}

\subsection{General equations}

Consider the theory describing the interaction of the Rarita-Schwinger field $\psi_{a i}$ with a Maxwellian electromagnetic field.
The Maxwell equations keep their form~\eqref{D_15}, and the Rarita-Schwinger equations describing massless spin-3/2  particles  are
\begin{equation}
	i \gamma^\mu \nabla_\mu \psi_{a i} = 0.
\label{RS_10}
\end{equation}
Here, the covariant derivative
$
	\nabla_\mu \psi_{a i} = \partial_\mu \psi_{a i} +
	\frac{1}{4} \omega_{b c \mu}
	\gamma^b_{i j} \gamma^c_{j k} \psi_{a k} -
	\omega^b_{\phantom{b} a \mu} \psi_{b i}-i e A_\mu \psi_{a i}
$  involves the spinor, $i, j, k$, tetrad,  $a, b, c$, and spacetime, $\mu$, indices, the four-current
$
	j^\mu =
	\epsilon^{abcd}\bar\psi_{ai}
	\left( \gamma_5\right)_{ij} \left( \gamma_b\right)_{jk}
	e^\mu_c \psi_{dk}
$
with $\epsilon^{abcd}$ being the completely antisymmetric Levi-Civita symbol.

The spinor Rarita-Schwinger field must also satisfy the following constraints~\cite{Redkov:2011ttu}
(hereafter, we drop the spinor index $i$ in $\psi_{a i}$ for simplicity):
\begin{align}
	\nabla^a \psi_a &= \frac{\kappa}{2} \gamma^a \psi_a ,
\label{RS_20}\\
	\left(\frac{1}{2} R_{a b}+i e F_{a b}\right) \gamma^a \psi^b +
	\left(
		\frac{\kappa^2}{2} - \frac{R}{12}-\frac{e}{3}F_{c d}\sigma^{c d}
	\right) \gamma^a \psi_a &= 0,
\label{RS_30}
\end{align}
where $\kappa$ is a numerical coefficient, $F_{ab}$ is the electromagnetic field tensor with tetrad indices, 
$R_{a b}$ and $R$ are the Ricci tensor and the scalar curvature, respectively. Taking into account the constraints~\eqref{RS_20} and \eqref{RS_30}, Eq.~\eqref{RS_10} can be recast in the following equivalent form:
\begin{equation}
	\epsilon^{\lambda \nu \rho \sigma} \gamma^5 \gamma_\nu
	\left(
		\nabla_\mu \psi_\nu - \nabla_\nu \psi_\mu
	\right)
	= 0,
\label{RS_40}
\end{equation}
where $\nabla_\mu \psi_\nu = e^a_{\phantom{a} \nu} \nabla_\mu \psi_a$;
in this form, it is used in SUGRA.

One of distinctive features of the Rarita-Schwinger equation is that the spinor $\psi_{a}$ has 16 components,
but the constraints~\eqref{RS_20} and \eqref{RS_30} reduce the number of independent components to 8. This means that this equation is overconstrained,
and the derivation of its solutions is hampered by this fact.

\subsection{Solutions with ``frozen'' electric and magnetic fields
}

In this case, there is only the Rarita-Schwinger equation~\eqref{RS_10} with the constraints~\eqref{RS_20} and \eqref{RS_30}.
Since we consider the case of a massless spinor field, we will seek a solution of Eq.~\eqref{RS_10} in $\mathbb{R} \times S^3$ spacetime with the metric~\eqref{D_20}
in the form
\begin{equation}
	\left( \psi_{a} \right)^T_{mn} = e^{-i \Omega t} e^{i n \chi} e^{i m \varphi}
	\begin{pmatrix}
		0	&	 0	&	 0	&	 0	&	 \\
		0	&	 0	&	 \Theta_{23}(\theta)	&	 \Theta_{24}(\theta)	&	 \\
		0	&	 0	&	 i \Theta_{33}(\theta)	&	 i \Theta_{34}(\theta)	&	 \\
		0	&	 0	&	 \Theta_{43}(\theta)	&	 \Theta_{44}(\theta)	&	 \\
	\end{pmatrix} =
	\left( \chi_{a} \right)^T_{mn} +
	\left( \xi_{a} \right)^T_{mn}.
\label{3_10}
\end{equation}
Here, we have split the Rarita-Schwinger spinor $\left( \psi_{a} \right)_{mn}$ into two Weyl spinors, one of which is set to be zero:
$\left( \chi_{a} \right)_{mn} = 0$. The integers $m$ and $n$ appearing here are quantum numbers, and
$(\dots)^T$ denotes the transpose for the spinor indices but not for the tetrad index $a$.

After substitution of the $\mathfrak{Ansatz}$~\eqref{3_10} in Eq.~\eqref{RS_10}, we get the following set of equations:
\begin{align}
	\Theta _{24}^\prime - \frac{\Theta _{33}}{2} -
	\frac{\Theta _{44}}{2} -
	\Theta _{23} \left( \frac{m}{\sin \theta} +
	n \cot \theta + \frac{r \Omega }{2} +
	\frac{1}{4}\right) +
	\Theta _{24} \left(\frac{\cot \theta}{2} - n\right) &= 0 ,
\label{3_20}\\
	\Theta _{34}^\prime -
	\frac{\Theta _{23}}{2} -
	\frac{\Theta _{44}}{2} -
	\Theta _{43} \cot \theta -
	\Theta _{33} \left(
		\frac{m}{\sin \theta} + n \cot \theta +
		\frac{r \Omega }{2} + \frac{1}{4}
	\right) +
	\Theta _{34} \left(\frac{\cot \theta}{2} - n\right)
	&= 0 ,
\label{3_30}\\
	\Theta _{44}^\prime +
	\frac{\Theta _{24}}{2} -
	\frac{\Theta _{34}}{2} -
	\Theta _{33} \cot \theta -
	\Theta _{43} \left(
		 \frac{m}{\sin \theta} + n \cot \theta +
		\frac{r \Omega }{2} + \frac{1}{4}
	\right) +
	\Theta _{44} \left(\frac{\cot \theta}{2} - n\right)	
	&= 0 ,
\label{3_40}\\
	- \Theta _{23}^\prime +
	\frac{\Theta _{34}}{2} +
	\frac{\Theta _{43}}{2} +
	\Theta _{24} \left(
		\frac{m}{\sin \theta} + n \cot \theta -
		\frac{r \Omega }{2} - \frac{1}{4}
	\right) -
	\Theta _{23} \left(\frac{\cot \theta}{2} + n\right)
	&= 0 ,
\label{3_50}\\
	- \Theta _{33}^\prime +
	\frac{\Theta _{24}}{2} -
	\frac{\Theta _{43}}{2} +
	\Theta _{44} \cot (\theta ) + \Theta _{34}
	\left(
		\frac{m}{\sin \theta} + n \cot \theta -
		\frac{r \Omega }{2} - \frac{1}{4}
	\right) -
	\Theta _{33} \left(\frac{\cot \theta}{2} + n\right)
	&= 0 ,
\label{3_60}\\
	- \Theta _{43}^\prime -
	\frac{\Theta _{23}}{2} -
	\frac{\Theta _{33}}{2} +
	\Theta _{34} \cot \theta +
	\Theta _{44} \left(
		\frac{m}{\sin \theta} + n \cot \theta -
		\frac{r \Omega }{2}-\frac{1}{4}
	\right) -
	\Theta _{43} \left(\frac{\cot \theta}{2} + n\right)
	&= 0 .
\label{3_70}
\end{align}

Let us now assume that the left-hand and right-hand sides of the constraint equation \eqref{RS_20} are equal to zero separately [below we will show that this assumption is valid for the solutions obtained,
see after Eq.~\eqref{3_190}].
This leads to the constraint	 $\gamma^a \psi_a = 0$, from which the following relations between the components of the spinor are found to be
\begin{equation}
	\Theta_{43} = -\Theta_{24} - \Theta_{34} , \quad \Theta_{44} = \Theta_{23} - \Theta_{33} .
\label{3_90}
\end{equation}
After substitution of~\eqref{3_90} in Eqs.~\eqref{3_20}-\eqref{3_70}, one can eliminate $\Theta_{23}^\prime,  \Theta_{33}^\prime, \Theta_{24}^\prime,  \Theta_{34}^\prime$
from Eqs.~\eqref{3_40} and \eqref{3_70} and get the following relations between the components of the spinor:
 \begin{align}
 	\Theta_{33} = & 2 \Theta _{23}\frac{
 		2 \frac{m}{\sin \theta} (2 r \Omega -1) +
 		8 n^2 + 2 n \cot \theta (2 r \Omega +1) +2 r \Omega -1
 	}
 	{16 n^2-(1 - 2 r \Omega )^2}
 \nonumber\\
 &
 	+2 \Theta _{24} \frac{
 		2 n (-4 \frac{m}{\sin \theta} + 2 r \Omega +1) +
 		\cot \theta \left(-8 n^2-2 r \Omega +1\right)
 	}
 	{16 n^2-(1-2 r \Omega )^2} ,
\label{3_110}
\end{align}
\begin{align}
 	\Theta_{34} = & - 2 \Theta _{23} \frac{
 		2 n (4 \frac{m}{\sin \theta} +2 r \Omega +1) +
 		\cot \theta \left(8 n^2 + 2 r \Omega -1\right)
 	}
 	{16 n^2-(1-2 r \Omega )^2}
\nonumber\\
 	&
 -	2 \Theta _{24}
 	\frac{
 		2 \frac{m}{\sin \theta} -
 		\frac{4 m r \Omega}{\sin \theta} +
 		8 n^2 - 2 n \cot \theta (2 r \Omega +1) +
 		2 r \Omega - 1
 	}{16 n^2-(1 - 2 r \Omega )^2} .
\label{3_120}
\end{align}
Substituting this in the initial set of equations, we find the following equations for the $\Theta_{23}$ and $\Theta_{24}$ components of the spinor:
\begin{align}
	\Theta _{23}^\prime +
		\Theta _{23} \left(\frac{\cot \theta}{2} + n \right) +
	\Theta _{24} \left(
		- \frac{m}{\sin \theta} - n \cot \theta +
		\frac{r \Omega }{2} + \frac{3}{4}
	\right)
	&= 0 ,
\nonumber\\
	\Theta _{24}^\prime + \Theta _{24}
		\left(\frac{\cot \theta}{2} - n \right) -
	\Theta _{23} \left(
		 \frac{m}{\sin \theta} + n \cot \theta +
		\frac{r \Omega}{2} + \frac{3}{4}
 		\right) &= 0 ,
\nonumber
\end{align}
where the eigenvalue $\Omega_n$ is defined as
\begin{equation}
	r \Omega_n = - \frac{3}{2} - 2 n .
\label{3_148}
\end{equation}
As in the case of the Dirac field [see after Eq.~\eqref{D_93}], one can see from this expression that, since the energy of the system $\Omega$ depends on $n$ only, the energy spectrum is degenerate with respect to $m$. 

We seek a solution of the above equations in the particular form
$$
	\Theta_{24} = \Theta_{23} = \Theta ,
$$
obtaining the equation
\begin{equation}
	\Theta^\prime + \Theta \left[
		\left( \frac{1}{2} - n \right) \cot \theta -
		\frac{m}{\sin \theta}
	\right] = 0 .
\label{3_149}
\end{equation}
Taking into account the expressions~\eqref{3_90}-\eqref{3_120}, we then have the following set of solutions:
\begin{align}
	\left( \Theta_{24}\right)_{mn} &=
	\left( \Theta_{23}\right)_{mn} = \Theta_{mn} =
	C  \tan^m \left(\frac{\theta }{2}\right)
	\sin^{n - \frac{1}{2}} \theta =
	\tilde C \sin^{n + m - \frac{1}{2}} \left(\frac{\theta }{2}\right)
	\cos^{n - m - \frac{1}{2}} \left(\frac{\theta }{2}\right) ,
\label{3_150}\\
	\left( \Theta_{33} \right)_{mn} &= C
	\sin^{n - \frac{1}{2}}(\theta )
	\tan^m \left(\frac{\theta }{2}\right)
	\left[
		(n - \frac{1}{2}) \cot \theta + \frac{m }{\sin \theta} + \frac{1}{2}
	\right] ,
\label{3_160}\\
	\left( \Theta_{34}\right)_{mn} &= C
	\sin^{n - \frac{1}{2}}(\theta )
	\tan^m \left(\frac{\theta }{2}\right)
	\left[
		(n - \frac{1}{2}) \cot \theta + \frac{m }{\sin \theta} - \frac{1}{2}
	\right] ,
\label{3_170}\\
	\left( \Theta_{43}\right)_{mn} &= - C
	\sin^{n - \frac{1}{2}}(\theta )
	\tan^m \left(\frac{\theta }{2}\right)
	\left[
		(n - \frac{1}{2}) \cot \theta + \frac{m }{\sin \theta} + \frac{1}{2}
	\right] ,
\label{3_180}\\
	\left( \Theta_{44}\right)_{mn} &= - C
	\sin^{n - \frac{1}{2}}(\theta )
	\tan^m\left(\frac{\theta }{2}\right)
	\left[
		(n - \frac{1}{2}) \cot \theta + \frac{m }{\sin \theta} - \frac{1}{2}
	\right] ,
\label{3_190}
\end{align}
where $C$ is an integration constant
and the functions appearing here are eigenfunctions of the Rarita-Schwinger equation. Substituting these solutions in Eqs.~\eqref{RS_20} and \eqref{RS_30}, one can show that they satisfy these constraints. Also, it can be shown that the solutions~\eqref{3_150}-\eqref{3_190} satisfy the Rarita-Schwinger equation~\eqref{RS_40} written in the form employed in SUGRA.

Thus the Weyl spinor $\left( \xi_{a} \right)_{mn}$ from~\eqref{3_10} can be recast in the form
\begin{equation}
	\left( \xi_{a} \right)^T_{mn} = C
	e^{-i \Omega t} e^{i n \chi} e^{i m \varphi}
	\sin^{\alpha} \left(\frac{\theta }{2}\right)
	\cos^\beta \left(\frac{\theta }{2}\right)
	\left\lbrace
	\left[
		\left( n - \frac{1}{2} \right) \cot \theta +
		\frac{m }{\sin \theta}
	\right]
	\left(
	\begin{array}{rr}
		0	&	0	\\
		0	&	0	\\
		1	&	1	\\
		- 1	&	- 1	
		\end{array}
	\right) + \frac{1}{2}
		\left(
			\begin{array}{rr}
				0	&	0	\\
				1	&	1 \\
				1	&	-1	\\
				-1	&	1
			\end{array}
		\right)
	\right\rbrace,
\label{3_220}
\end{equation}
where $\alpha = n + m - \frac{1}{2},
\beta = n - m - \frac{1}{2}
$.

Also, there exists one more solution, for which
$$
	r \Omega = - \frac{3}{2} + 2 n , \quad	\Theta_{24} = - \Theta_{23}=\Theta .
$$
This leads to the replacement of $m, n$ by $- m, - n$ in Eq.~\eqref{3_149} and to the corresponding replacements in the expressions~\eqref{3_150}-\eqref{3_190}.

\subsubsection{Properties of the solution}

We demand that the solution~\eqref{3_220} would be integrated by quadrature  
$$
	\int \left( \xi^\dagger_a \right) _{mn}
	\left( \xi_a \right)_{mn} dV < \infty , \quad
	\text{no summing over } a .
$$
Then we have the following restrictions on the possible values of the quantum numbers $m$ and $n$:
\begin{equation}
	n \geqslant 2 , \quad
	\left| m \right| \leqslant n - \frac{1}{2} .
\label{mn_lims}
\end{equation}
This in turn leads to the following inequality for the energy coming from the energy spectrum~\eqref{3_148}:
$$
	r \Omega_n \leqslant - \frac{11}{2} .
$$

For the Rarita-Schwinger field, the symmetrized energy-momentum tensor with tetrad indices is calculated by the formula
$$
	T_{a b} = \frac{1}{2}
	\left(
		\epsilon^{m \phantom{a} c d}_{\phantom{m} a}
			e_c^{\phantom{c} \rho} \bar{\psi}_m
			\gamma_5 \gamma_b
			\nabla_\rho \psi_d + 	
		\epsilon^{m \phantom{b} c d}_{\phantom{m} b}
		e_c^{\phantom{c} \rho} \bar{\psi}_m
		\gamma_5 \gamma_a
		\nabla_\rho \psi_d
	\right) + c.c.
$$
Substituting here the solutions~\eqref{3_150}-\eqref{3_190}, we have
\begin{equation}
\begin{split}
	T_{a b} = &C^2
	\frac{ (4 n+3) }{r}
	\sin ^{2 n - 1} \theta
	\sin^{2 m}\left(\frac{\theta }{2}\right)
	\cos^{-2 m} \left(\frac{\theta }{2}\right)
\nonumber\\
	&
\times	\left\lbrace
	2 \left[
			\frac{m}{\sin \theta} + \left(n - \frac{1}{2} \right) \cot \theta
	\right]^2
	\begin{pmatrix}
		1	&	0	&	0	&	0	\\
		0	&	1	&	0	&	1	\\
		0	&	0	&	0	&	0	\\
		0	&	1	&	0	&	0	\\
	\end{pmatrix} +
		\begin{pmatrix}
			\frac{3}{2}	&	0	&	0	&	0	\\
			0	&	- \frac{1}{2}	&	0	&	0	\\
			0	&	0	&	1	&	0	\\
			0	&	0	&	0	&	1	\\
		\end{pmatrix}
	\right\rbrace .
\end{split}
\end{equation}


The spin density for the Rarita-Schwinger field is defined as
$$
	S_{abc} = \frac{i}{4} \left[
		\bar \psi_b \gamma_a \psi_c -
		\bar \psi_c \gamma_a \psi_b + \eta_{ab} \left(
				\bar \psi_c \gamma^d \psi_d - \bar \psi_d \gamma^d \psi_c
		\right) - \eta_{ac} \left(
		\bar \psi_b \gamma^d \psi_d - \bar \psi_d \gamma^d \psi_b
		\right)
	\right] .
$$
For the solutions~\eqref{3_150}-\eqref{3_190}, there are the following nonzero components:
\begin{align}
	S_{023} =& - S_{032} = S_{123} = - S_{132} = S_{324} =
	- S_{342} = A
	\left[
		\frac{m}{\sin \theta } +
		\left( n - \frac{1}{2} \right) \cot \theta
	\right],
\nonumber\\
	- S_{213} =& = S_{231} = S_{312} = - S_{321} = A
	\frac{\tan \theta}{2} ,
\nonumber\\
	S_{023} =& - S_{032} = S_{123} = - S_{132} = A
	\tan \theta \left\lbrace
		\frac{1}{4} +
		\left[
			\frac{m}{\sin \theta} +
			\left( n - \frac{1}{2}\right) \cot \theta
		\right]^2
	\right\rbrace ,
\nonumber
\end{align}
where
$$
A = 	\frac{C^2 (4 n+3)}{r}
	\sin^{2 n - 1} \theta
	\sin^{2 m}\left(\frac{\theta }{2}\right)
	\cos^{-2 m}\left(\frac{\theta }{2}\right).
$$

\section{Conclusions}

Summarizing the results obtained,
\begin{itemize}
\item We have shown that in $\mathbb{R} \times S^3$ spacetime the Dirac and  Rarita-Schwinger equations allow separation of variables.
\item In the absence of electric and magnetic fields, these equations have regular solutions.
\item In the presence of an electromagnetic field the set of Dirac-Maxwell equations has self-consistent solutions with nonzero electric and magnetic fields.
\item In Dirac-Maxwell theory, the current of the Dirac field is related to the Hopf invariant.
\item Both for the Dirac field and for the Rarita-Schwinger field, the energy-momentum and spin density tensors have been calculated.
\end{itemize}

Let us also point out the following features of the Dirac-Maxwell system  studied in the present paper. The set of Dirac-Maxwell equations permits a solution for which the spinor and magnetic fields are regular at the point $\theta = \pi$,
whereas the electric field is singular at that point. The peculiarity of the situation is that the singularity of the
electromagnetic potentials $A_{t, \chi, \varphi}$ does not lead to a singularity of the spinor.

In conclusion, the solutions obtained can be of considerable interest, since the solution with the Dirac field is related to the topological invariant of the Hopf fibration.
In turn, the solution with the Rarita-Schwinger spinor can be of some interest in respect of its use in describing the fermionic sector of  SUGRA.
Note also that the solutions found can be regarded as an analogue of plane waves in Minkowski space, since there are no external fields in these solutions.
The main difference from the plane waves consists in the fact that the solutions obtained here form a discrete spectrum numbered by integers $m$ and $n$.

\section*{Acknowledgments}
We gratefully acknowledge support provided by Grant No.~BR05236494
in Fundamental Research in Natural Sciences by the Ministry of Education and Science of the Republic of Kazakhstan. We are also grateful to the Research Group Linkage Programme of the Alexander von Humboldt Foundation for the support of this research.

\appendix

\section{Hopf fibration}
\label{Hopf_bundle}

The standard three-dimensional sphere $S^3$ is a set of points in $\mathbb{R}^4$  that satisfy the equation
$$
X^2 + Y^2 + Z^2 + W^2 = r^2,
$$
where $r$ is the radius of the sphere and $X, Y, Z, W$ are coordinates of the enveloping four-dimensional space in which the three-dimensional sphere is embedded.

The fibration $h: S^3 \rightarrow S^2$ is defined by the Hopf mapping
$$
h \left( X, Y, Z, W \right) = \left\{
X^2 + Y^2 - Z^2 - W^2,
2 \left( X W + Y Z \right) ,
2 \left( Y W - X Z \right)
\right\}.
$$
The Hopf coordinates  $0 \leq \theta \leq \pi$, $0 \leq \chi, \varphi \leq 2 \pi$ on a three-dimensional sphere are given in the form
\begin{eqnarray}
	X &=& r \cos \left( \frac{\varphi + \chi}{2} \right) \sin \frac{\theta}{2},
\quad
	Y = r \sin \left( \frac{\varphi + \chi}{2} \right) \sin \frac{\theta}{2},
\label{A_40}\nonumber\\
	Z &=& r \cos \left( \frac{\varphi - \chi}{2} \right) \cos \frac{\theta}{2},
\quad
	W = r \sin \left( \frac{\varphi - \chi}{2} \right) \cos \frac{\theta}{2}.
\label{A_60}\nonumber
\end{eqnarray}
On the Hopf fibration, one can introduce the so-called Hopf invariant, which is defined as
$$
	H = \frac{1}{V} \int \Upsilon \wedge d \Upsilon ,
$$
where $\Upsilon$ is some 1-form and the volume of a three-dimensional Hopf sphere is
$V = \int \sin \theta d \chi d \theta d \varphi$. On the Hopf fibration, the 1-form
 $$
	\Upsilon = d \chi - \cos\theta d \varphi .
$$
For this case, the Hopf invariant
$
  H = 1 .
$

\end{document}